\documentstyle[12pt]{article}
\begin{flushright}
{\normalsize McGill/93-36\\
September 1993 \\}
\end{flushright}
\centerline{\Large LONGEVITY OF}
\centerline{\Large QUARK-GLUON PLASMA AND THE MIXED PHASE} 
\centerline{\Large FROM INTENSITY INTERFEROMETRY}
\centerline{\Large OF HIGH ENERGY PHOTONS}
\centerline{\bf Dinesh Kumar Srivastava$^*$ and Charles Gale }
\centerline  {\it Department of Physics, McGill University}
\centerline{\it 3600 University Street, Montr\'{e}al, H3A 2T8, Canada}
\date{}
\begin{document}
\begin{center}
Abstract\\
\end{center}      

Two-photon intensity interferometry is shown to provide an accurate
measurement of lifetime of  quark-gluon plasma
 created in ultra-
relativistic heavy ion collisions via the difference of outward and sideward
correlation radii.
Under the  assumption of a longitudinal, boost invariant expansion of the
plasma, we obtain analytical expressions for the correlations from the
quark-gluon plasma phase. A $3+1$ dimensional expansion of the plasma along with
a first order phase transition to hadrons is next considered, and,
 leads to a source with two characteristic
lifetimes, one for the quark-gluon plasma phase, and the other for the longer
lived mixed phase. This may even help us to {\em experimentally} determine the
 order of the phase transition.

\vskip 50pt
\noindent
$^*$ {\small Permanent address: {\it Variable Energy Cyclotron Center, 1/AF,
Bidhan Nagar, Calcutta 700 064, India}}
\vfill \eject

It is believed by now that energy densities large enough to encompass
 a  chiral symmetry restoring/ deconfining
phase transition may be attained in 
collisions involving large nuclei at the Relativistic Heavy Ion Collider or the
Large Hadron Collider. A considerable progress has  been made in
 understanding the early stages ("the first 2--3 fm/c") of the collision in
terms of the parton cascade model \cite{Klaus} or the simpler "hot gluon
scenario" \cite{hotglue}.
The plasma created in such collisions will expand, cool, and get dilute; paving
the way for hadronization.
The resulting hadrons would  decouple from the interacting system possibly 
over  a very short time interval at  freeze out.
Thus as the hadrons appear only in the final stages of such
collisions, their correlations will mainly carry
information about the late dilute stage of the collision, not about
the early dense stage. 

In contrast to hadrons, photons
are produced throughout the space-time evolution of the reaction, and
suffer essentially no interactions with the surrounding medium once
they are produced.  This has led to studies \cite{DJ1,DJ2,DJ3,DJ4} 
investigating the
utility and the fidelity of 
intensity interferometry of
photons with large transverse momenta.
These ought to be created primarily in the quark-gluon plasma because
that is when the temperature available to produce them is greatest.

In the present work we demonstrate the ability of photon interferometry
to provide a most valuable information about the lifetime of the
QGP and the mixed phase in terms of the difference of the outward and sideward
correlation radii. In order to clarify these concepts, first of all, we obtain
analytic expression for the correlation function for photons radiated
from the QGP, by using
a boost invariant longitudinal expansion \cite{BJ}
 and approximating the time dependence
of the source function as an exponential with a time scale, $\tau_0$.

Results for a full $3+1$ dimensional expansion \cite{hydro,Jane}
 with a first order phase
transition are then presented, without any approximation. Essentially two
lifetimes are seen to emerge, one for the QGP phase of a shorter
duration and the other for the mixed phase of a much larger duration. It is
felt that this result may even help us {\em experimentally} 
determine the much debated order of the
phase transition since for a second order phase transition the mixed phase
will be absent.

The correlation between two photons with momenta ${\bf k}_1$ and
${\bf k}_2$ and the same helicity is
\begin{equation}
C({\bf k}_1,{\bf k}_2) = \frac{P({\bf k}_1,{\bf k}_2)}
{P({\bf k}_1)P({\bf k}_2)}
~~,{\rm with}~
P({\bf k}) = \int d^4x \frac{dN(x,{\bf k})}{d^4xd^3k}
\end{equation}
and
\begin{equation}
P({\bf k}_1,{\bf k}_2) = \int d^4x_1 d^4x_2 \frac{dN(x_1,{\bf k}_1)}
{d^4x_1d^3k_1} \frac{dN(x_2,{\bf k}_2)}{d^4x_2d^3k_2}
[1 + \cos(\Delta k \cdot \Delta x)]
\end{equation}
The rates per unit volume for producing a
photon 
 with momentum ${\bf k}$ at the space-time point $x$,
 from  QGP and hadronic matter at a given temperature  $T$ 
 are known to be nearly equal \cite{us} and given by
\begin{equation}
E\frac{dN}{d^4xd^3k} = K\,T^2\,\ln \left( \frac{2.9}{g^2} \frac{E}{T}
+ 1 \right) \exp (-E/T)
\end{equation}
where $E$ is the photon energy, $g$ is the QCD coupling constant,
and $K$ is a constant which is irrelevant for the correlation function.

The correlation function $C({\bf k}_1,{\bf k}_2)$ is routinely 
examined \cite{Sc,Be,Fe}
in terms of the longitudinal $(q_L)$, sideward $(q_{\rm side})$ and
 outward $(q_{\rm out})$
momentum differences of the two particles. It can be argued that
the corresponding correlation radii, $R_L$, $R_{\rm side}$, and $R_{\rm out}$,
provide the longitudinal, transverse, and (lifetime + transverse) measurements
of the source. Thus the duration of particle emission is expected to be
given by $(R_{\rm out}-R_{\rm side})$.

If the four-momentum of the $i$-th photon is written as
\begin{equation}
k_i^\mu=(k_{\rm iT}\,\cosh y_i,{\bf k}_i)~~,
~~{\bf k}_i=(k_{\rm iT}\,\cos \psi_i,k_{\rm iT}\,\sin\psi_i,k_{\rm iT}\,
\sinh y_i)~,
\end{equation}
where $k_{\rm iT}$ is the transverse momentum, $y_i$ is the rapidity and
$\psi_i$ is the azimuthal angle, we have \cite{DJ2},
\begin{equation}
{\bf q}_T={\bf k}_{\rm 1T}-{\bf k}_{\rm 2T}~~,~~
{\bf K}_T=({\bf k}_{\rm 1T}+{\bf k}_{\rm 2T})/2~,
\end{equation}
\begin{equation}
q_L=k_{\rm 1z}-k_{\rm 2z}~~,~~q_{\rm out}=\frac{{\bf q}_T\cdot{\bf K}_T}
{K_T}~~,~{\rm and}~ q_{\rm side}=|{\bf q}_T-q_{\rm out}\frac{{\bf K}_T}{K_T}|
\end{equation}

As a first step, we illustrate the physics to be measured by this correlation 
function with a simple dynamical model of the collision: Bjorken hydrodynamics
\cite{BJ}.  This model has been used to estimate both lepton pair
\cite{KKMM} and photon production \cite{Dinesh}.  The necessary developments
for photon interferometry have also been given \cite{DJ1,DJ2}, and we have,
$P({\bf k}_1,{\bf k}_2) = P_1P_2 + P_{c1}P_{c2} + P_{s1}P_{s2}$
where
\begin{equation}
P_i = P({\bf k}_i) = \pi R^2 K \int d\tau \, \tau \,
\sqrt{\frac{2\pi T}{k_{iT}}} \, T^2 \, \ln \left( \frac{2.9}{g^2}
\frac{k_{iT}}{T} + 1 \right) \exp(-k_{iT}/T)~~,~
\end{equation}
\begin{eqnarray}
P_{ci} &=& \pi R^2 K \int d\tau \, \tau \,
\sqrt{\frac{2\pi T}{k_{iT}}} \, T^2 \, \ln \left( \frac{2.9}{g^2}
\frac{k_{iT}}{T} + 1 \right) \exp(-k_{iT}/T) \nonumber \\
&& \left[ \frac{2J_1(q_TR)}{q_TR} \right] \,
\cos[(\Delta E \, \cosh y_i - q_L \, \sinh y_i) \tau].
\end{eqnarray}
Here $J_1$ is the Bessel function, $R$ is the radius of the identical
nuclei undergoing a central collision, and
$\Delta E = k_{1T} \, \cosh y_1 - k_{2T} \, \cosh y_2$.
The $P_{si}$ are the same as the $P_{ci}$ with the substitution
of a sine for the cosine.

The time dependence of the temperature for a plasma expanding according
to Bjorken hydrodynamics is well known \cite{DJ1,DJ2}.
In Fig.1, we have shown the (proper) time
 dependence of the photon emitting source (essentially the integrand of eq.(7),
without the slowly varying logarithmic term),
for photons having transverse momenta of 1 GeV and rapidity $y=0$.
We choose $T_c$ = 160 MeV, and at RHIC, $~~~T_i$=532 MeV at
$\tau_i = 1/3T_i$ = 0.124 fm/c \cite{Klaus,Jane,KMS} with
$T_f$ = 140 MeV for a collision involving two lead nuclei.  This model and
these numbers are used for illustrative purposes.

We see the three distinct time scales for the source function. During the
QGP phase, one may possibly 
approximate the source function as $\exp(-\tau/\tau_0)$. Next
we have a long lived mixed phase and a hadronic phase of a brief duration. 

{\it If we concentrate on the QGP phase only}, and replace the integration
over $[\tau_i,\tau_Q]$ by $[0,\infty]$ in the above,
 the correlation function can be written as,
\begin{equation}
C=1+\left[\frac{2J_1(q_TR)}{q_TR}\right]^2\times
\frac{1+\alpha_1\alpha_2\tau_0^2}{(1+\alpha_1^2\tau_0^2)(1+\alpha_2^2\tau_0^2)}
~,
\end{equation}
\begin{equation}
 \alpha_1=k_{\rm 1T}\,\cosh (y_1-y_2)-k_{\rm 2T}~~,~~
 \alpha_2=k_{\rm 1T}-k_{\rm 2T}\,\cosh (y_1-y_2)~~.
\end{equation}

After some algebra, we can show that the correlation function for
QGP phase alone, under the Bjorken hydrodynamic expansion can be decomposed
for $y_2=0$,
as
\begin{eqnarray}
C(q_{\rm side},~ q_{\rm out}=0,~q_L=0)&=&1+\left[\frac{2J_1(q_{\rm side}
R)}{q_{\rm side}R}\right]^2\\
C(q_{\rm side}=0,~ q_{\rm out},~q_L=0)&=&1+\left[\frac{2J_1(q_{\rm out}
R)}{q_{\rm out}R}\right]^2\times\frac{1}{1+q_{\rm out}^2\tau_0^2}\\
C(q_{\rm side}=0,~ q_{\rm out}=0,~q_L)&=&1+\frac{1-\Delta E^2\tau_0^2}
{\left(1+\Delta E^2\tau_0^2\right)^2}
\end{eqnarray}

It is interesting to recall that a somewhat similar expression
was obtained by Kopylov
and Podgoretski \cite{KP} for emissions from a static spherical
source of radius R and time scale $\tau_0$.
Recalling the approximation $(2J_1(x)/x)^2\approx e^{-x^2/4}$ suggested by
Zajc \cite{Zajc}, and further approximating $(1+x)^{-1} \approx \exp (-x) $
 for small $x$, we see that for the case of longitudinal expansion,
 $R_{\rm side}$ is the transverse radius of the system R, and
 {\mbox{$R_{\rm out}\approx\sqrt{R_{\rm side}^2+4\tau_0^2}$},
 confirming our earlier 
surmise that the difference of $R_{\rm out}$ and $R_{\rm side}$ 
gives us information about the duration
of the particle emitting process.

 The result for longitudinal correlation
is very interesting indeed, since $q_L$ does not explicitly appear in this
expression (eq.13). The longitudinal correlation function
 is seen to be determined by
the energy difference of the photons and the time scale $\tau_0$.
 A closer scrutiny however reveals that
 this aspect is a consequence of the boost invariance
of the flow, assumed in Bjorken hydrodynamics. In a boost invariant
scenario for the flow, correlation between photons emitted from $(t_1,z_1)$ and
$(t_2,z_2)$ should be decided by the difference of the 
 proper times $\tau_1=\sqrt{t_1^2-z_1^2}$ and $~~\tau_2=\sqrt{t_2^2-z_2^2}$,
 whose analogue is energy difference.

Next we obtain results for a $3+1$ dimensional expansion \cite{DJ2,Jane}
of the plasma, undergoing a first order phase transition, for a collision
involving two lead nuclei at energies reached at RHIC. The results for
outward and sideward correlations with the longitudinal momentum difference
$q_L$ taken as zero are given in Fig.2. Further we have taken, $q_{\rm out}$
as zero while evaluating the sideward correlation and, similarly, taken 
$q_{\rm side}$ as zero, while evaluating the outward correlation. 
Photons having transverse momenta around 1 GeV are chosen 
for these model calculations. For the ease of extraction of the correlation
radii we have plotted $\ln (C-1)$ vs. $q_{\rm side}^2$ or $q_{\rm out}^2$.
While the sideward correlation is seen to be characterized by only one
scale, the outward correlation apparently has two scales. Thus the 
outward correlation is better represented as
\begin{equation}
C(q_{\rm out},~q_{\rm side}=0,~q_L=0)\approx 1+a\exp\left[-q_{\rm out}
^2R_{1({\rm out})}^2/4\right]+
b\exp\left[-q_{\rm out}^2R_{2({\rm out})}^2/4\right]+...
\end{equation}
 Even though
this two scale behaviour was present for outward correlation function in
earlier model calculations as well \cite{DJ2}, its significance was not
fully realized there. The results given here were also obtained
by a more accurate determination of the multidimensional integrals
involved for the correlations for a $3+1$ dimensional flow.
 This exercise leads to
$R_{\rm side}=$ 8.6 fm, and  two scales for $R_{\rm out}$, as 22 fm and 9.8 fm,
respectively. If we take the cue from the relations (11) and (12) above,
 we get two time scales
 $\tau_0=\frac{1}{2}\sqrt{R_{\rm out}^2-R_{\rm side}^2}$
from this analysis, one equal to about 10 fm/c, and the other equal to about 
2.4 fm/c. It is satisfying to note that photon interferometry provides results
consistent with the hydrodynamic description \cite{Jane} 
 employed for the description of
the evolution of the system. Similar results were obtained by 
analysing the predictions 
 for $k_T$ values of 2 and 3 GeV. Obviously the smaller lifetime 
corresponds to the longevity of
the plasma phase, and the larger time is for the mixed phase. This was again 
checked by analysing the correlations for the QGP and the mixed phases
separately, as in Ref. \cite{DJ2}. Please note that hadronic phase makes 
only a very small contribution to thermal photons due to much 
cooler temperatures during that phase. It is  of interest to recall that
two-proton interferometric study \cite{Go}
 of heavy-ion collisions at the energy of a
 few hundred MeV/nucleon,  led to two time-scales, one from slow particle
evaporation, and the other from nonequilibrium particle emission. Incidentally
the pion-interferometric study of the ultrarelativistic nuclear
collisions by the NA35 collaboration \cite{Fe} leads to a
difference of $R_{\rm out}$ and $R_{\rm side}$  consistent
with zero. This implies that the pions do not probe the
 history of evolution of the system.

Thus we see that intensity interferometry of high energy photons is able to
provide the lifetime of the mixed and the QGP phase. This
information can be of vital importance, as the lifetime of the mixed phase
is crucially dependent on the rate of nucleation and 
the rate of cooling \cite {Be,Laszlo}.
In addition, the lifetimes of the phases constructed in this model study
are decided by the equations of state for the hadronic matter and the QGP.
If the QGP phase transition is second order in nature
the mixed phase will be absent.

Before proceeding  we would like to remark that,
thus far we have assumed that the polarization of each photon is
measured.  If a polarization average is taken instead, then
the factor $1 + \cos(\Delta k \cdot \Delta x)$ in eq. (2)
gets replaced by $1 + \frac{1}{2}\cos(\Delta k \cdot \Delta x)$.
This means that $C \rightarrow \frac{3}{2}$ at zero relative
momentum rather than 2.  Otherwise the behavior of the correlation
function is unchanged.

A systematic analysis of the feasibility of these studies with a proper 
accounting of energy resolution of the detectors etc. is underway. Looking at
Fig.2, we realize that if the energy resolution for the photons is much worse
 than 1--2\%, it may not be possible to see at least the
 outward correlations for the larger correlation radius.
The longitudinal correlations presented in \cite{DJ1,DJ2,DJ3,DJ4} earlier
have however much larger scales in terms of $q_L$ (smaller in terms of
$R_L$) and also show two scales for the time evolution \cite{DJ3}, which should
make their study interesting.

 The ominous
question of background from misidentified decay photons from $\pi^0$ decay
can not be answered easily. It is  felt that the $p_T$ spectrum for $\pi^0$
for $AA$ system can not be accurately obtained by extrapolating the
results from $pp$ studies due to degradation, rescattering and absorption of
the (mini)jets in nuclear medium. 
The hydrodynamics employed earlier to evaluate the flow can not
be used with confidence for evaluating the pion spectra, due to
the uncertainty associated with the criterion for freeze out
for a transversely expanding system,
and the large contributions of (mini)jets at larger transverse momenta.
It is felt that for a system undergoing a transverse expansion, the pions
may remain in thermal contact till the temperatures are much lower,
 (mean free paths much  larger), where the simple equation of state
utilized in these studies may not be valid any more.

  The background problem may be less severe
for photons having larger $k_T$, (see \cite {DJ1,DJ2,DJ3} for more discussion).
However, it is also felt that detection of thermal photons having larger
 transverse momenta is linked with detection of $\pi^0$ having large
transverse momenta, through their decay photons. Their precise detection would
require a detector of fine granularity, placed at a large distance,
 as proposed by the PHENIX collaboration, say.
 A simulation study by using a substantial number of events from FRITIOF
or VENUS is perhaps the best one can do at the moment, and such a study is under
progress.

In brief we have tried to focus upon a unique and valuable information,
about the longevity of QGP and mixed phase
obtainable from intensity interferometry of high energy photons which may
be produced in reasonable abundance in collisions involving heavy nuclei at
RHIC and at LHC. This information along with the space-time history of
the evolution of the system which can be studied using photon interferometry
makes it a very powerful and useful concept indeed. The extent to which
 these aspects
will  be realized by the innovative ingenuity of experimental efforts remains
 to be seen.
\section*{Acknowledgements}
We are  grateful  to Prof. P. Braun-Munzinger for his comments on 
the feasibility of photon interferometry at RHIC energies 
which led to this investigation.
Very helpful discussions and communications with
J. Barrette, S. Das Gupta, and H. Gutbrod,
are thankfully acknowledged. One of us (DKS) would like to thank
members of Department of Physics at McGill University in general and 
 Prof. S. Das Gupta in particular for a very warm hospitality during his stay
in Montr\'{e}al where this work was completed. We both would like to
acknowledge the warm hospitality of the Institute for Theoretical Physics at
the University of California in Santa Barbara where parts of this work were
done. This work was funded in part by the Natural Sciences and Engineering
Research Council of Canada, in part by le Fond FCAR du Qu\'ebec, and in part by
NSF under grant number PHY89--04035.

\newpage
\section*{Figure Captions}

Figure 1:  The time evolution of the source function emitting thermal
photons at $y=0$. A plasma created at initial temperature ($T_i$) of 532 MeV, at
the proper time $1/3T_0$ and undergoing expansion according to
Bjorken hydrodynamics is considered. The system is in QGP phase up to $\tau_Q$,
in the mixed phase during $\tau_Q$ to $\tau_H$, and undergoes freeze out
at $\tau_F$. 

\vspace{.25in}

\noindent Figure 2:  Sideward and outward correlation function for
thermal photons having transverse momenta around 1 GeV. A $3+1$ dimensional
expansion of the system is considered along with a first order phase
transition. We take $q_L=0$ and consider collision of two lead nuclei at
RHIC energies. Additionally we have
 $q_{\rm out}=0$ for the sideward
correlation and $q_{\rm side}=0$ for the outward correlation.
We have plotted $\ln (C-1)$  vs. $q_{\rm side}^2$ or $q_{\rm out}^2$ for the
ease of extraction of correlation radii. 
Similar results, with two scales for the outward correlation were seen for
other values of transverse momenta.

\end{document}